\documentstyle[12pt,axodraw,epsf]{article}
\begin{document}
\baselineskip 0.6cm
\newcommand{\gsim}{ \mathop{}_{\textstyle \sim}^{\textstyle >} }
\newcommand{\lsim}{ \mathop{}_{\textstyle \sim}^{\textstyle <} }
\newcommand{\vev}[1]{ \left\langle {#1} \right\rangle }
\newcommand{\bra}[1]{ \langle {#1} | }
\newcommand{\ket}[1]{ | {#1} \rangle }
\newcommand{\EV}{ {\rm eV} }
\newcommand{\KEV}{ {\rm keV} }
\newcommand{\MEV}{ {\rm MeV} }
\newcommand{\GEV}{ {\rm GeV} }
\newcommand{\TEV}{ {\rm TeV} }
\newcommand{\tr}{ {\rm tr} }
\newcommand{\diag}{ {\rm diag} }
\def\Spin{\mathop{\rm Spin}}
\def\SO{\mathop{\rm SO}}
\def\O{\mathop{\rm O}}
\def\SU{\mathop{\rm SU}}
\def\U{\mathop{\rm U}}
\def\Sp{\mathop{\rm Sp}}
\def\SL{\mathop{\rm SL}}


\begin{titlepage}

\begin{flushright}
UT-985
\end{flushright}

\vskip 2cm
\begin{center}
{\large \bf  Higher Dimensional Supersymmetry as an Origin \\}
{\large \bf   of the Three Families for Quarks and Leptons   }

\vskip 1.2cm
T.~Watari$^a$ and T.~Yanagida$^{a,b}$

\vskip 0.4cm
$^{a}$ {\it Department of Physics, University of Tokyo, \\
         Tokyo 113-0033, Japan}\\
$^{b}$ {\it Research Center for the Early Universe, University of Tokyo,\\
         Tokyo 113-0033, Japan}

\vskip 1.5cm
\abstract{In a (0,1) supersymmetric (SUSY) six-dimensional gauge theory, 
a gauge fermion gives rise to box anomalies.
These anomalies are completely 
canceled by assuming a vector multiplet of (1,1) SUSY.
With a ${\bf T}^2/{\bf Z}_3$ orbifold compactification of the extra 
two-dimensional space, 
the theory provides three chiral multiplets and three
equivalent fixed points.
We regard them as the origin of the three families of quarks and leptons.
Quasi anarchy structure in the SU(5)-{\bf 5}$^*$ sector 
and hence the bilarge mixing in the neutrino oscillation are 
explained quite naturally in this framework.
We also discuss a family symmetry as a remnant of the higher-dimensional 
R symmetry.}

\end{center}
\end{titlepage}


\section{Introduction}

The triplicate family structure of quarks and leptons 
is one of the most mysterious features in particle physics, 
and it is expected to be an important indication 
of more fundamental physics beyond the standard model. 
In the four-dimensional spacetime all gauge anomalies are canceled
within one family and there is no necessity to introduce other two
families in nature.
However, the anomaly cancellation in higher-dimensional theories imposes 
further nontrivial conditions on the theories, and it sometimes requires
multiplication of massless particles.

In this letter, we show that the triplicate family structure arises
naturally from a six-dimensional supersymmetric (SUSY) gauge theory. 
We assume a SUSY SO(10) gauge theory and put the vector multiplet 
of SO(10) in the six-dimensional bulk.
The anomalies in the six-dimensional spacetime are completely canceled
out by introducing the (1,1) SUSY vector multiplet in the bulk.
We adopt a ${\bf T}^2/{\bf Z}_3$ orbifold to reduce the theory to a
four-dimensional theory with ${\cal N}$=1 SUSY.
We find that the SO(10) is broken down to SU(5)$\times$U(1)$_5$ and 
three families of ${\bf 10}$'s of the SU(5) remain massless after the
orbifolding.
The [SU(5)]$^3$ anomalies are localized at the three independent fixed
points, which are canceled by introducing one ${\bf 5}^*$ 
at the each fixed point ({\it i.e.,} three ${\bf 5}^*$'s).
The three massless ${\bf 10}$'s come from three SO(10)-{\bf adjoint}
${\cal N}$=1 chiral multiplets $\Sigma,\Sigma'$ and $\Sigma''$ in the
(1,1) SUSY vector multiplet.
Thus, we consider that the (1,1) SUSY in the six-dimensional spacetime 
is the origin of three families of quarks and leptons\footnote{
Bosonic components of these three $\Sigma,\Sigma'$ and $\Sigma''$ multiplets
may be regarded as six independent polarizations (45,67 and 89
directions) of a ten-dimensional vector field, since the (1,1) SUSY 
vector multiplet in the six dimensions can be obtained through a
compactification of ${\cal N}$=1 vector multiplet of the ten-dimensional 
spacetime.
In this sense, the triplicate family structure may originate from the extra 
six space dimensions.}.

We have an SU(2)$_{{\bf 4}_{+}}$ family symmetry which is a subgroup 
of the R symmetry of the (1,1) SUSY.
The SU(5)-${\bf 10}$'s transform as ${\bf 1}+{\bf 2}$ under this 
SU(2)$_{{\bf 4}_{+}}$ symmetry while the SU(5)-${\bf 5}^*$'s are singlets.
The breaking of the SU(2)$_{{\bf 4}_{+}}$ symmetry leads to 
the observed mass hierarchies of the quarks and leptons and
the CKM mixing angles.
The bilarge mixing in the neutrino sector\cite{SK} is naturally explained,
since the ${\bf 5}^*$ are all equivalent to each other\cite{anarchy}.

\section{A SUSY SO(10) gauge theory in the six-dimensional spacetime}

Let us consider a SUSY SO(10) gauge theory in the six-dimensional
spacetime. 
The vector multiplet of the (0,1) SUSY gives rise to irreducible box
anomalies\cite{FK,AW}, and these anomalies must be canceled 
by introducing suitable SO(10)-charged hyper multiplets in the
bulk\footnote{It is pointed out in Ref.\cite{FK} that the anomaly is
finite and unique despite the nonrenormalizability of the theory. 
T.Y. thanks M.~Shifman for raising this issue at ``Peccei's Fest''.}.
The simplest and the most beautiful way to do this is to
introduce an (1,1) SUSY vector multiplet as a whole.
Once we assume the (1,1) SUSY in the six-dimensional spacetime, 
then not only the irreducible box anomalies but also reducible anomalies
vanish. 
It is not necessary to resort to the Green-Schwartz mechanism\cite{GS,genGS6} 
or to introducing extra particles.
There is no global anomaly either\cite{global1,global2}.
If one takes a torus (${\bf T}^2$) compactification of the extra two
space dimensions $x_4$ and $x_5$, we obtain 
a Lagrangian for the Kaluza-Klein zero modes which has ${\cal N}$=4 SUSY
in the four-dimensional spacetime.
The Kaluza-Klein zero modes form an ${\cal N}$=4 SO(10) vector multiplet,
${\cal W}_\alpha,\Sigma,\Sigma'$ and $\Sigma''$, 
where the ${\cal W}_\alpha$ is the field strength tensor 
(${\cal N}$=1 vector multiplet) and the rests are SO(10)-{\bf adjoint} 
${\cal N}$=1 chiral multiplets. 
The two scalar components of the $\Sigma$ are the two independent 
polarization modes of the six-dimensional SO(10) vector field 
in the fourth and fifth dimensions.

The R symmetry of the (1,1) SUSY in the six-dimensional spacetime 
is SU(2)$_{{\bf 4}_{-}} \times$ SU(2)$_{{\bf 4}_{+}}$\footnote{For
conventions adopted in this paper, see \cite{IWY} or 
the appendix of this paper.}$\!\!\!\!$.
These SU(2)$_{{\bf 4}_{-}}$ and SU(2)$_{{\bf 4}_{+}}$ are R symmetries
that are associated to the six-dimensional SUSY charges,
\begin{equation}
 {\cal Q}_{{\bf 4}_{-}}^{(6)} = \left(\begin{array}{cc}
			   & - {\cal Q}_\alpha^{(4) 2} \\
              -\bar{{\cal Q}}^{(4) \dot{\alpha}}_1&
				\end{array}\right) 
\quad {\rm ~and~} \quad
 {\cal Q}_{{\bf 4}_{+}}^{(6)} = \left(\begin{array}{cc}
		{\cal Q}^{(4) 4}_{\alpha}& \\
       & \bar{{\cal Q}}^{(4) \dot{\alpha}}_3
				      \end{array}\right),
\end{equation}
which belong to ${\bf 4}_{-}$ and ${\bf 4}_{+}$ spinor representations
of the SO(5,1), respectively. 
After a Kaluza-Klein reduction, the six-dimensional SUSY charges are
decomposed into four four-dimensional SUSY charges, 
${\cal Q}^{(4)1,2,3,4}_\alpha$, and not only the 
SU(2)$_{{\bf 4}_{-}} \times $ SU(2)$_{{\bf 4}_{+}}$ R symmetry
but the rotational symmetry SO(2)$_{45}$ of the extra space dimensions  
is also regarded as a subgroup of the
SU(4)$_{\rm R}$ R symmetry of the ${\cal N}$=4 SUSY in four-dimensional
spacetime ({\it i.e.,} SO(2)$_{45} \times $SU(2)$_{{\bf 4}_{-}} \times $
SU(2)$_{{\bf 4}_{+}}\subset $SU(4)$_{\rm R}$). 
The four (four-dimensional) SUSY charges ${\cal Q}^{(4)1,2,3,4}_\alpha$ 
transform as a ${\bf 4}$ representation of the SU(4)$_{\rm R}$, 
four fermions in the four chiral multiplets (of Kaluza-Klein zero modes), 
${\cal W}_\alpha,\Sigma,\Sigma'$ and $\Sigma''$, as a 
${\bf 4}^*$ representation, 
and six real scalars in the $\Sigma,\Sigma'$ and $\Sigma''$ as 
a ${\bf 4}^* \wedge {\bf 4}^* \simeq {\bf 6}$, respectively. 
In particular, we note for the later purpose that 
the SU(2)$_{{\bf 4}_{+}}$ commutes with the ${\cal N}$=1 SUSY transformation 
generated by ${\cal Q}^{(4)1}$ 
and that the two ${\cal N}$=1 chiral multiplets
$\Sigma'$ and $\Sigma''$ form a doublet of this SU(2)$_{{\bf 4}_+}$ 
symmetry (see {\it e.g.,} \cite{IWY} or the appendix).

\begin{figure}[ht]
\begin{center}
\begin{picture}(200,200)(-100,-100)
\Line(-100,52)(100,52)
\Line(-100,0)(100,0)
\Line(-100,-52)(100,-52)
%
%
 \Vertex(-60,35){2} \Vertex(-60,0){2} 
 \Vertex(  0,35){2} \Vertex(  0,0){2} \Vertex(  0,-35){2}
                    \Vertex( 60,0){2} \Vertex( 60,-35){2}
%
%
\Vertex(-90,52){2}                                        
\Vertex(-30,52){2} \Vertex(-30,17){2} \Vertex(-30,-17){2} \Vertex(-30,-52){2}
\Vertex( 30,52){2} \Vertex( 30,17){2} \Vertex( 30,-17){2} \Vertex( 30,-52){2}
                                                          \Vertex( 90,-52){2}
\Line(-105,-26)(-75,-78)
\Line(-105,78)(-15,-78)
\Line(-45,78)(45,-78)
\Line(15,78)(105,-78)
\Line(75,78)(105,26)
%

%
\Text(-90,52)[lb]{3}  \Text(-30,52)[lb]{3}  \Text( 30,52)[lb]{3}
\Text(-60,35)[lb]{2}  \Text(  0,35)[lb]{2}
\Text(-30,17)[lb]{1}  \Text( 30,17)[lb]{1} 
\Text(-60,0)[lb]{3}   \Text(  0,0)[lb]{3}   \Text( 60,0)[lb]{3}
\Text(-30,-17)[lb]{2} \Text( 30,-17)[lb]{2}
\Text(  0,-35)[lb]{1} \Text( 60,-35)[lb]{1}
\Text(-30,-52)[lb]{3} \Text( 30,-52)[lb]{3} \Text( 90,-52)[lb]{2}
\end{picture}
\end{center}
\caption{A picture of the ${\bf T}^2/{\bf Z}_3\vev{\sigma}$ geometry is
given. 
Unit cell of the ${\bf T}^2$ torus is described by parallel lines
and three fixed points labeled by 1,2,3 are given on it.
One can see that all three fixed points are equivalent to each other.}
\label{fig:orbifold}
\end{figure}
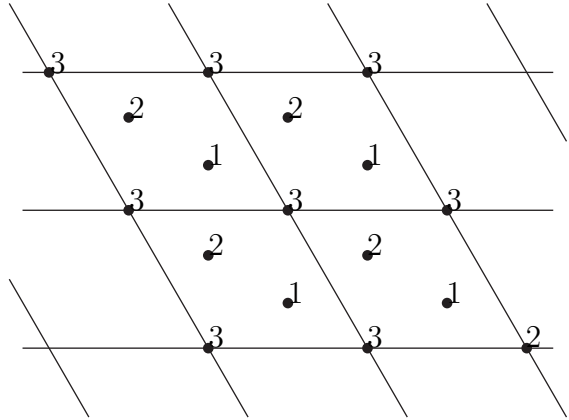

Now we compactify the (1,1) SUSY SO(10) gauge theory on the ${\bf
T}^2/{\bf Z}_3\vev{\sigma}$ geometry rather than on the ${\bf T}^2$
(see Fig.\ref{fig:orbifold}) to obtain a four-dimensional theory 
with only ${\cal N}$=1 SUSY\footnote{
It is argued in Ref.\cite{BGT} that the number of family can be
constrained by considering the gauge anomalies in six-dimensional 
field theories. 
However, ``the number of family in the six-dimensional spacetime''
which they discuss has, in principle, no direct connection to 
the number of family as they admit by themselves.
It is only after giving a definite way and/or principle to obtain 
chiral fermions in the four-dimensional spacetime that the use 
of higher-dimensional spacetime makes sense
in discussing the number of family and family structure.
In fact, one obtains any number of families in their approach.}.
The generator $\sigma$ rotates the extra two-dimensional space
by angle $-(2/3) 2\pi$:
\begin{equation}
\sigma : (x_4 + i x_5) \mapsto \omega^{-2} (x_4 + i x_5),
\label{eq:rotation}
\end{equation}
where $\omega \equiv e^{2\pi i / 3}$.
Some of Kaluza-Klein zero modes are always projected out 
from the Hilbert space of the theory on the orbifold ${\bf
T}^2/{\bf Z}_3\vev{\sigma}$.
We take, here, the orbifold projection conditions as follows\cite{orbifold};
\begin{eqnarray}
{\cal W}_\alpha &=& \quad \;\; \gamma_\sigma {\cal W}_\alpha \gamma^{-1}_\sigma, \label{eq:OPC1}\\
\Sigma   \;     &=& \omega^{-2} \gamma_\sigma \Sigma \;\; \gamma^{-1}_\sigma,
\label{eq:OPC2}\\
\Sigma'  \,     &=& \omega \quad \gamma_\sigma \Sigma' \; \gamma^{-1}_\sigma,
\label{eq:OPC3}\\
\Sigma''        &=& \omega \quad \gamma_\sigma \Sigma'' \; \gamma^{-1}_\sigma,
\label{eq:OPC4} 
\end{eqnarray}
where we have taken an SU(4)$_{\rm R}$-twist as  
$\diag(1,\omega^2,\omega^{-1},\omega^{-1}) \in$ SU(4)$_{\rm R}$. 
The gauge-twisting matrix $\gamma_\sigma$ associated with the
generator $\sigma$ is 
\begin{equation}
\gamma_\sigma = \diag(\overbrace{\omega,\cdots,\omega}^5,
               \overbrace{\omega^{-1},\cdots,\omega^{-1}}^5) \in {\rm SO(10)}
\end{equation}
in the Cartan-diagonal base.  
Note that the SO(2)$_{45}$ rotation, 
$\diag(\omega,\omega,\omega^{-1},\omega^{-1}) \in$ SU(4)$_{\rm R}$, given 
in Eq.(\ref{eq:rotation}) is accompanied by a twist of the internal symmetry, 
$\diag(\omega^{-1},\omega,1,1) \in$ SU(2)$_{{\bf 4}_{-}}$, so that 
the combined SU(4)$_{\rm R}$-twist, 
$\diag(1,\omega^2,\omega^{-1},\omega^{-1})$, 
belongs to an SU(3) subgroup of the SU(4)$_{\rm R}$; 
the ${\cal N}$=1 SUSY survives when and only when the SU(4)$_{\rm
R}$-twist belongs to the SU(3) subgroup of which the ${\cal
Q}^{(4)1}_{\alpha}$ is singlet\cite{CHSW,orbifold}.

The SO(10) gauge symmetry is now broken down to SU(5)$\times$U(1)$_5$, and
the massless particles remaining in the Hilbert space are 
the SU(5)$\times$U(1)$_5$ ${\cal N}$=1 vector multiplets and 
three SU(5)-${\bf 10}$ ${\cal N}$=1 chiral multiplets.
The vector multiplets arise from the ${\cal W}_\alpha$ and 
the three {\bf 10}'s from $\Sigma,\Sigma'$ and $\Sigma''$.
We identify these three {\bf 10}'s with those of 
quarks and leptons\footnote{A coset space SO(10)/(SU(5) $\times$
U(1)$_5$) contains SU(5)-{\bf 10}.
Possible connection between this fact and the origin of the SU(5)-{\bf
10} of quarks and leptons was pointed out long time ago 
in \cite{BLPY,KY}.}\footnote{
Ref.\cite{LPT} also obtains three {\bf 10}'s 
from SO(11) ten-dimensional ${\cal N}$=1 vector multiplet. They
considered the Type I string theory on a ${\bf T}^6/{\bf Z}_3$
orientifold, and three families of SU(5)-${\bf 5}^*$ also survives 
the orbifold projection there. However, this model is 
not acceptable as a realistic model because too rapid proton decay 
is inevitable through dimension-four operators.}.
Therefore, the origin of the three families is 
the (1,1) SUSY in the six-dimensional spacetime ({\it i.e.,} 
${\cal N}$=4 SUSY in the four-dimensional spacetime).

Because there are three families of ${\bf 10}$'s in the bulk, 
we have [SU(5)]$^3$ anomaly.
Such an anomaly localizes only at fixed points of the orbifold\cite{ACG}.
In the present ${\bf T}^2/{\bf Z}_3\vev{\sigma}$ orbifold  
the anomaly distributes at the three fixed points with the same amount 
since they are all equivalent to each other\footnote{
We can also confirm this distribution by an explicit calculation.
The distribution function is given by
\begin{equation}
A({\bf y}) = 3 \sum_{{\bf k}}
( |\psi^{{\bf Z}_3}_{{\bf 10},{\bf k}}({\bf y})|^2
 - |\psi^{{\bf Z}_3}_{{\bf 10}^*,{\bf k}}({\bf y})|^2 )
\end{equation}
where ${\bf k}$ runs for all Kaluza-Klein momenta, ${\bf y}$ is the
coordinate of the torus ${\bf T}^2$ and the wave functions are   
\begin{equation}
\psi^{{\bf Z}_3}_{{\bf 10}^{(*)},{\bf k}} = 
\frac{1}{3 \sqrt{V} }\sum_{n = 0}^{2} 
   \gamma^n_{{\bf 10}^{(*)}} e^{i {\bf k} \cdot (\sigma^n \cdot {\bf y})}
\qquad {\rm where~} 
\gamma_{{\bf 10}}=1 {\rm ~and~} \gamma_{{\bf 10}^*}= \omega^{-1},
\end{equation}
and $V$ is the volume of the torus ${\bf T}^2$.}(see Fig.\ref{fig:orbifold}).
Therefore, the simplest way to cancel these anomalies
is to introduce a ${\bf 5}^*$ at each fixed point 
({\it i.e.,} three ${\bf 5}^*$'s as a whole).
We identify these ${\bf 5}^*$'s with the three families of ${\bf 5}^*$'s 
of quarks and leptons.
The charges of these ${\bf 5}^*$'s under the surviving 
U(1)$_5$ gauge symmetry is still arbitrary, and hence
the mixed anomalies U(1)$_5 \cdot$[SU(5)]$^2$ does not vanish in general.
However, if the U(1)$_5$ charge of the
${\bf 5}^*$'s are (-3) times those of the ${\bf 10}$'s, then, this
anomaly is automatically canceled at each fixed point. 
Other mixed anomalies U(1)$_5 \cdot$[gravity]$^2$ and [U(1)$_5$]$^3$ 
can be canceled simultaneously by introducing a right-handed neutrino 
at each fixed point.
This extra U(1)$_5$ gauge symmetry, which is usually called 
as the fiveness, is a linear combination of 
the U(1)$_{\rm Y}$ and U(1)$_{\rm B-L}$, and the small neutrino masses 
are naturally obtained by the see-saw mechanism\cite{GRS-Y} 
when the U(1)$_{\rm B-L}$ is spontaneously broken.
Even if the U(1)$_5$ charge of the {\bf 5}$^*$'s does not satisfy the
above relation, all three anomalies discussed above can be canceled by
invoking a generalized Green-Schwarz mechanism\cite{GS,genGS4} at each
fixed point. 

If one starts with an E$_6$ vector multiplet of the (1,1) SUSY in the
six-dimensional spacetime, then three families of SO(10)-{\bf 16} 
survives the orbifold projection on the ${\bf T}^2/{\bf Z}_3$\cite{BLPY,KY}.
In this case, there is no [SO(10)]$^3$ anomaly at each fixed point.

\section{Discussion}

We discuss, in this section, phenomenological consequences of the R symmetry 
SO(2)$_{45} \times$ SU(2)$_{{\bf 4}_{-}} \times$ SU(2)$_{{\bf 4}_{+}}$.
Since the three families of ${\bf 10}$'s originate from a single 
vector multiplet of the (1,1) SUSY, some part of the R symmetry 
becomes a inter-family symmetry and some part becomes a low-energy 
R symmetry of the four-dimensional ${\cal N}$=1 SUSY.

\subsection{SU(2)$_{{\bf 4}_{+}}$ family symmetry}

The orbifold projection conditions Eqs.(\ref{eq:OPC1}-\ref{eq:OPC4}) 
do not violate the SU(2)$_{{\bf 4}_+}$ symmetry, and hence
we assume this symmetry to be preserved even in the theory 
on the ${\bf T}^2/{\bf Z}_3$ orbifold.
This SU(2)$_{{\bf 4}_{+}}$ symmetry is a pure family
symmetry in the sense that the Grassmann coordinates of
the ${\cal N}$=1 superspace do not rotate under this symmetry.
The three families of ${\bf 10}$'s transform as ${\bf 1}+{\bf 2}$
(which we denote as ${\bf 10}_3+{\bf 10}_a|_{a=1,2}$) 
under this SU(2) 
family symmetry.
On the other hand, the ${\bf 5}^*$'s are all singlets of 
the family symmetry SU(2)$_{{\bf 4}_{+}}$, since the {\bf 5}$^*$'s 
localize at different fixed points while the SU(2)$_{{\bf 4}_{+}}$
is an internal symmetry and it does not exchange fields on separated
points of spacetime. 
The anarchy structure in the neutrino sector\cite{anarchy} 
is naturally expected in this framework, since 
there is no distinction between the three ${\bf 5}^*$'s.

We introduce Higgs multiplets, $H({\bf 5})$ and $\bar{H}({\bf 5}^*)$, 
at one of the fixed points.
Given this  situation, we have to consider the orbifold geometry 
whose length scale is of order of the fundamental scale, 
because otherwise the Yukawa couplings would be highly suppressed for 
the two families of ${\bf 5}^*$'s which do not localize 
at the same fixed point of the Higgs multiplets, and the resulting
mass spectra of the down-type quarks and charged leptons 
would be unrealistic. 
Therefore, the three fixed points should be close to each other, 
and no suppression of interaction is expected between fields that localize
at different fixed points.
Thus, there is no essential difference between the three ${\bf 5}^*$'s,
and the quasi anarchy structure is still expected in the {\bf
5}$^*$'s.

As long as the SU(2)$_{{\bf 4}_{+}}$ family symmetry is unbroken, 
only one family of quarks and charged leptons acquire their masses:
\begin{equation}
 W = y \; {\bf 10}_3 {\bf 10}_3 H({\bf 5}) + 
     y_i \; {\bf 5}^*_i \cdot {\bf 10}_3 \cdot \bar{H}({\bf 5}^*) 
\qquad \qquad i=1,2,3. 
\end{equation}
On the other hand, Majorana neutrino masses are allowed by 
the family symmetry for all three families :
\begin{equation}
 W = \frac{\kappa_{ij}}{M_R}{\bf 5}^*_i H({\bf 5}) {\bf 5}^*_j H({\bf 5})
\qquad \qquad i,j = 1,2,3.
\end{equation}
Coefficients $y, y_i$ and $\kappa_{ij}$ are expected to be of order 1. 
The above result explains why the masses are large for 
the quarks and charged leptons in the third family and why the mixings 
among families are large in the neutrino sector\cite{SK}.

Now we have to introduce small breaking of the SU(2)$_{{\bf 4}_{+}}$ 
symmetry so that the the first and the second families of quarks and 
leptons are able to obtain their non-zero masses.
Let us suppose that the SU(2)$_{{\bf 4}_{+}}$ breaking is implemented
through two doublets,
\begin{equation}
\phi^a = \left( \begin{array}{c} 0 \\ \epsilon \end{array} \right)
\qquad \qquad \qquad 
\tilde{\phi}^a = \left( \begin{array}{c} 
                    \epsilon' \\ \epsilon'' \end{array}\right),
\label{eq:SU(2)-breaking}
\end{equation}
where we assume $\epsilon',\epsilon'' \lsim {\cal O}(\epsilon)$. 
Then, the mass matrices are roughly given by
\begin{equation}
m_u \sim \left(\begin{array}{ccc}
	   \epsilon^{'2} & \epsilon \epsilon' & \epsilon' \\
	   \epsilon \epsilon' & \epsilon^2 & \epsilon \\
	   \epsilon' & \epsilon & 1 \\
		  \end{array}\right),
m_{d,e} \sim \left(\begin{array}{ccc}
	   \epsilon' & \epsilon & 1 \\
	   \epsilon' & \epsilon & 1 \\
	   \epsilon' & \epsilon & 1 \\
		  \end{array}\right),
m_{\nu} \sim \left(\begin{array}{ccc}
	   1 & 1 & 1 \\
	   1 & 1 & 1 \\
	   1 & 1 & 1 \\
		  \end{array}\right),
\end{equation}
which are obtained from additional superpotential,
\begin{equation}
W =  {\bf 10}_3 ({\bf 10}\cdot \Phi) H  
  + ({\bf 10}\cdot \Phi)({\bf 10} \cdot \Phi) H
  + y'_i \; {\bf 5}^*_i \cdot ({\bf 10} \cdot \Phi) \cdot \bar{H},
\end{equation}
where $\Phi$ represents $\phi$ and $\tilde{\phi}$. 
In particular, the empirical relation $m_s/m_b \sim m_\mu/m_\tau \sim 
\epsilon$, $|V_{cb}| \sim \epsilon$ and $m_{c}/m_{t} \sim \epsilon^2$ is 
obtained.
The bilarge mixing in the neutrino sector is also a preferable
consequence of the anarchy structure of the ${\bf 5}^*$'s.
If one assumes $\epsilon' \sim \epsilon^2$, then the mass matrices are
similar to that in the Froggatt-Nielsen framework\cite{FN}, and their
phenomenological success is discussed in various references\cite{texture,HM}.
Note that we have neglected the effects of the GUT breaking.
One can also understand the violation of the SU(5) GUT relation 
in the masses of the first and the second families, if one takes 
account of contributions that involve GUT breaking vacuum-expectation
values. 

\subsection{Low-energy R symmetry}

It is clear that the SO(2)$_{45} \times $SU(2)$_{{\bf 4}_{-}}$ subgroup 
is not preserved in the orbifold projection conditions 
Eqs.(\ref{eq:OPC1}-\ref{eq:OPC4}). 
Only the SO(2)$_{45}$ and the Cartan part U(1)$_{{\bf 4}_{-}}$ 
of the SU(2)$_{{\bf 4}_{-}}$ can be preserved.
A suitable linear combination of these SO(2)$_{45}$ and 
the U(1)$_{{\bf 4}_{-}}$ yields a U(1) R symmetry\footnote{Here, 
a term ``R symmetry'' is used in its narrow sense: a symmetry
that rotates the Grassmann coordinates of the ${\cal N}$=1 SUSY.}
which might be relevant in the low-energy physics. 
This U(1) R symmetry is specified so that the three families of 
${\bf 10}$'s have the same charge: 2/3.
We assume that the other independent linear combination of 
the two U(1) symmetries is broken down (otherwise 
necessary Yukawa couplings would be forbidden).
The charge assignment of the U(1) R symmetry is not determined for 
the three ${\bf 5}^*$'s or Higgs multiplets, since their origin 
is not clear. 
However, those charges are fixed by phenomenological requirements 
that the U(1) R symmetry allows up-type and down/charged-lepton-type
Yukawa couplings and Majorana neutrino masses: that is, 1/3 for 
${\bf 5}^*$, 2/3 for $H({\bf 5})$ and 1 for $\bar{H}({\bf 5}^*)$.
Then, as a consequence, the notorious dimension-four proton decay
operators $W = {\bf 5}^* \cdot {\bf 10} \cdot {\bf 5}^*$ and an enormous
mass-term for the Higgs multiplets are forbidden\footnote{
The color-triplet components of the $H({\bf 5})$ and $\bar{H}({\bf
5}^*)$ can receive the GUT-breaking mass term keeping 
this U(1) R symmetry\cite{ss}.}.

Anisotropy of the orbifold geometry ${\bf T}^2/{\bf Z}_3$, however,
might lead to further breaking of this low-energy R symmetry\footnote{
If the SU(2)$_{{\bf 4}_+}$-symmetry breakings given in 
Eq.(\ref{eq:SU(2)-breaking}) were charged also under this symmetry, 
then, the symmetry would be broken further. 
However, we assume that the SU(2)$_{{\bf 4}_+}$-symmetry breaking 
is not charged either under the SU(2)$_{{\bf 4}_-}$ or under 
the SO(2)$_{45}$.
This assumption does not lead to any appearent contradicitions, since we
do not know the origin of the breaking Eq.(\ref{eq:SU(2)-breaking}).};
only a discrete subgroup of the SO(2)$_{45}$ symmetry is left unbroken.
We show, in the following, that even if there are operators that violate
continuous U(1) R symmetry due to this anisotropy,
the U(1) R symmetry is preserved by\footnote{
There are two ways in describing a discrete subgroup of a U(1) symmetry:
one is the ``${\bf Z}_N$ subgroup'' and the other is the ``U(1) mod
charge $Q$''. 
The former is used when one can take all U(1) charges to be
integers, while the latter is used when there is a canonical
normalization of the U(1) charges.
$N = Q$ when all charges are integers in their canonical normalization, 
but it is not always the case.
In the present case, the U(1) R symmetry have a canonical
normalization ({\it i.e.,} the ${\cal N}$=1 Grassmann coordinates 
have charge 1 and a superpotential has charge 2), and 
some fields have fractional charges in this normalization. 
Thus, we take the latter description ``U(1) R symmetry mod charge Q'' 
in the text.
If one would rescale the R charges by multiplying 3 to make all the
charges integral (then the superpotential has R charge 6), then 
``the U(1) R mod charge 2'' could be referred to as ``${\bf Z}_6$ R symmetry''.
Proton decay operators ${\bf 5}^* \cdot {\bf 10}\cdot {\bf 5}^*$ and 
${\bf 5}^* {\bf 10}{\bf 10}{\bf 10}$ have R charge 4 and 7,
respectively, and the doublet-Higgs mass term 5 
in this normalization.} mod charge 2.
This is because when the ${\bf 10}$'s transform as
${\bf 10}(\theta) \rightarrow e^{-i \alpha 2/3}{\bf 10}(e^{i\alpha}\theta)$,
the U(1) R symmetry,
$\diag(e^{i\alpha/3},e^{i\alpha/3},e^{-i\alpha/3},e^{-i\alpha/3})\cdot 
\diag(e^{i\alpha2/3},e^{-i\alpha 2/3},1,1) \in $ SO(2)$_{45} \times $ 
SU(2)$_{{\bf 4}_{-}} \subset $SU(4)$_{\rm R}$, rotates 
the extra-dimensional space ${\bf T}^2$ by 
$(x_4 + i x_5) \rightarrow e^{-i\alpha 2/3} (x_4 + i x_5)$; The orbifold 
geometry we consider has a discrete rotational symmetry by angle $2\pi/3$,
and this means that the R symmetry is preserved for $\alpha \in \pi
{\bf Z}$. 
The R symmetry preserved by mod 2 is sufficient to forbid 
the dimension-four proton decay operators and the Higgs mass term, 
since the R charges of these operators are $4/3 \neq 2$ (mod 2) 
and $5/3 \neq 2$ (mod 2), respectively.

\section*{Acknowledgments}

We are very grateful to Martin Schmaltz for valuable comments.
T.W. thank the Japan Society for the Promotion of Science for 
financial support.
This work was partially supported by ``Priority Area: Supersymmetry and
Unified Theory of Elementary Particles (\# 707)'' (T.Y.).

\appendix
\section{SUSY and R Symmetry in the Six-dimensional Spacetime}

\subsubsection*{Chirality}
Spinor representations of the SO(5,1) is ${\bf 4}_+ \oplus{\bf 4}_-$.
Both ${\bf 4}_+$ and ${\bf 4}_-$ representations have four
complex-valued components.
These four-component spinors are regarded as Dirac spinors of the
SO(3,1), or in other words, are comprised of (1/2,0)-representation 
(left-handed) and (0,1/2)-representation (right-handed) 
two-component Weyl spinors.
Although ${\bf 4}_{\pm}$ representations are also sometimes referred to 
as left-handed and right-handed spinors, we do not use these
terminologies in order not to make confusion with the same `left- 
and right-' for spinors of SO(3,1).

Complex conjugate of the ${\bf 4}_+$ spinor is isomorphic to itself,
and so is the ${\bf 4}_-$. That is,
\begin{equation}
{\bf 4}_+^* \simeq {\bf 4}_{+}  \qquad \qquad {\bf 4}_-^* \simeq {\bf 4}_-.
\end{equation}
This is in contrast to the fact that the complex conjugate of 
the (1/2,0)-representation is the (0,1/2)-representation and vice versa
in the four-dimensional spacetime.
Therefore, in the six-dimensional spacetime, spinor fields
and their complex conjugates (in other words, fermionic states and 
their CPT conjugates) belong to the same spinor representation 
of the SO(5,1) ({\it i.e.,} ${\bf 4}_+$ or ${\bf 4}_-$), {\it that is,} 
they have a definite chirality.

Mass partners of ${\bf 4}_+$-spinor fields(states) are ${\bf 4}_-$ 
and vice versa. This is also in contrast with the fact that 
the mass partner of a (1/2,0)-spinor is again an (1/2,0)-spinor 
in the four-dimensional spacetime.
In particular, box anomalies from a ${\bf 4}_-$-state are opposite to
those of a ${\bf 4}_+$-state with otherwise the same representation.

\subsubsection*{SUSY and R symmetry}
Each SUSY charge in the six-dimensional spacetime, which is 
a spinor of the SO(5,1), has definite chirality. 
This is why the SUSY of the six-dimensional field
theory is characterized by a pair of non-negative integers 
$({\cal N}_+,{\cal N}_-)$; ${\cal N}_+$ and ${\cal N}_-$ are the number of 
SUSY charges that transform as ${\bf 4}_+$ and ${\bf 4}_-$, respectively.
In gauge theories with (0,1) SUSY, the SUSY charge and
fermions in hyper multiplets are ${\bf 4}_-$-spinors while the parameter 
of the SUSY transformation and fermions in vector multiplets are
${\bf 4}_+$-spinors.
In gauge theories with (1,1) SUSY there are SUSY charges in ${\bf 4}_+$
and ${\bf 4}_-$, transformation parameters in ${\bf 4}_-$ and ${\bf
4}_+$, and gauge fermions in ${\bf 4}_-$ and ${\bf 4}_+$.
In particular, all box anomalies are canceled within a single 
vector multiplet in (1,1) SUSY gauge theories, since vector multiplets
contain both ${\bf 4}_+$ and ${\bf 4}_-$ gauge fermions.

One can think of an SU(2) transformation that exchanges a SUSY charge
and its complex conjugate, since the complex conjugate belong to the
same spinor representation as the original one\footnote{
For more detail, see section III and the appendix of \cite{IWY}.}. 
This internal symmetry that acts on SUSY charges is an R
symmetry of the six-dimensional SUSY theories.
There are two independent SU(2) transformations in the (1,1) SUSY
theories; one (which we denote as SU(2)$_{{\bf 4}_+}$) acts on the 
SUSY charge in ${\bf 4}_+$-spinor, and the other (SU(2)$_{{\bf 4}_-}$)
on the SUSY charges in ${\bf 4}_-$-spinor.
Thus, the R symmetry of the (1,1) SUSY theories is SU(2)$_{{\bf 4}_-} 
\times $ SU(2)$_{{\bf 4}_+}$.

\subsubsection*{Kaluza-Klein reduction}
Let us take a toroidal compactification and consider 
an effective theory of Kaluza-Klein zero-modes. 
(0,1) SUSY gauge theories become ${\cal N}$=2 SUSY theories, since
the SUSY charge ${\cal Q}^{(6)}_{{\bf 4}_-}$ of six-dimensional 
theories consists of two independent SUSY charges 
${\cal Q}^{(4) 1}_\alpha$ and ${\cal Q}^{(4) 2}_\alpha$ of
four-dimensional theories.
(1,1)-SUSY theories become ${\cal N}$=4 SUSY theories with four 
SUSY charges ${\cal Q}^{(4)1 \cdots 4}_\alpha$.
Gauge fermions also decompose into four Weyl spinors
$\chi_{\alpha,1\cdots 4}$; the ${\bf 4}_+$ gauge fermion into
the $\chi_{1,2}$ and the ${\bf 4}_-$ gauge fermion into the $\chi_{3,4}$.
The ${\cal Q}^{(4)1,2}$ and $\chi_{1,2}$ are doublets of the SU(2)$_{{\bf
4}_-}$ R symmetry, and ${\cal Q}^{(4)3,4}$ and $\chi_{3,4}$ are doublets
of the SU(2)$_{{\bf 4}_+}$.

Rotational symmetry of the two extra-dimensional space SO(2)$_{45}$ that 
was originally a subgroup of the Lorentz symmetry SO(5,1) is now
regarded as an internal symmetry. This SO(2)$_{45}$ is also an R symmetry, 
since the SUSY charges transform non-trivially under the SO(5,1) and
hence under the SO(2)$_{45}$.

It is useful to invoke SU(4) notation in summarizing relation between
various R symmetries (SU(2)$_{{\bf 4}_{-}} \times $SU(2)$_{{\bf
4}_{+}}$ and the SO(2)$_{45}$) and their action on various fields.
SU(4) is the maximal R symmetry of the ${\cal N}$=4 SUSY gauge
theories on the four-dimensional spacetime and the above R symmetries
are regarded as subgroups of the SU(4)$_{\rm R}$ symmetry.
SUSY generators ${\cal Q}^{(4)a}_\alpha$ ($a=1,\cdots ,4$) are ${\bf 4}$
of the SU(4)$_{\rm R}$ symmetry, SUSY transformation parameters and gauge
fermions $\chi_{\alpha, a}$ are ${\bf 4}^*$, and 4 $\times$ 4 2nd rank
antisymmetric tensor (${\bf 4}^* \wedge {\bf 4}^*$) $\varphi_{ab}$ with 
a reality condition $\varphi_{ab}\epsilon^{abcd}/2 = \varphi^{* \,cd}$ are 
six scalars of the ${\cal N}$=4 multiplet, among which $\varphi_{12}
\propto (A_4 + i A_5)$, the two independent polarizations of the vector
field in the extra two space directions.
The SU(2)$_{{\bf 4}_-} \times $SU(2)$_{{\bf 4}_+}$ R symmetry is
included in the SU(4)$_{\rm R}$ as
\begin{equation}
 \left( \begin{array}{cc}
  \SU(2)_{{\bf 4}_-} & \\ & \SU(2)_{{\bf 4}_+} 
	\end{array}\right) \subset \SU(4)_{\rm R}
\end{equation}
in the 4 $\times$ 4 fundamental representation. The SO(2)$_{45}$
subgroup is included in the SU(4)$_{\rm R}$ as 
\begin{equation}
\diag (e^{i\alpha},e^{i\alpha},e^{-i\alpha},e^{-i\alpha}) 
       \subset \SU(4)_{\rm R},
\end{equation}
which corresponds to the rotation of the 4-th and 5-th plane by
\begin{equation}
   (x_4 + ix_5) \rightarrow e^{-2i\alpha}  (x_4 + ix_5) \qquad 
   (A_4 + iA_5) \rightarrow e^{-2i\alpha}  (A_4 + iA_5).
\end{equation}

${\cal N}$=1 SUSY generated by ${\cal Q}^{(4)1}$ survives the orbifolding, 
as long as the SU(4)$_{\rm R}$-twist (see the text) of the orbifold
projection is included in the lower right SU(3) subgroup of the
SU(4)$_{\rm R}$.
Under this ${\cal N}$=1 SUSY, $\chi_1$ is the gaugino, and the $\chi_2$
is the SUSY partner of the $\varphi_{12}\propto(A_4 + iA_5) $.
It will be clear that the SU(2)$_{{\bf 4}_+}$ now acts like an ordinary
(non-R) symmetry in the four-dimensional effective theory with ${\cal
N}$=1 SUSY, since it keeps the $\chi_1$ invariant.


%

\end{document}